\documentclass{article}

\usepackage{amsmath, amsthm, amssymb}
\usepackage{graphicx}
\usepackage{graphics}
\usepackage{subfig}
\usepackage[a4paper]{geometry}
\begin{document}

\title{Simplified solutions of the Cox-Thompson inverse scattering method at fixed energy}
\author{Tam\'as P\'almai$^1$, Mikl\'os Horv\'ath$^2$ and Barnab\'as Apagyi$^1$\\
\ \\
$^1$Department of Theoretical Physics\\
$^2$Department of Mathematical Analysis\\ Budapest University of
Technology and Economics\\ H-1111 Budapest, Hungary\\
\ \\
apagyi@phy.bme.hu}

\maketitle

\begin{abstract}
Simplified solutions of the Cox-Thompson inverse quantum
scattering method at fixed energy are derived if a finite number of partial waves
with only even or odd angular momenta are contributing to the
scattering process. Based on new formulas various approximate
methods are introduced which prove applicable also to the generic
scattering events.
\end{abstract}

\section{Introduction}
The Cox-Thompson (CT) inversion method  \cite{Cox1970} has a
number of useful properties. As input it requires a finite set
consisting of $N$  phase shifts. As output it produces  potentials
which possess non-vanishing first momentum, $\int{rV(r)
{\mathrm{d}}r \ne 0}$, and finite values at the origin, $\vert
V(0)\vert< \infty$ \cite{Apagyi2003}. These properties make the CT
method an attractive constructive  procedure which can   be
applied to experimental phase shifts in order to   determine
interaction potentials between composite quantum systems. However,
the CT method leads to the solution of equations which are of
nonlinear character, involving the inverse of  a matrix containing
the unknown quantities themselves. The solution of the CT
equations is therefore numerically difficult to perform and,
sometimes, only successful with deployment of sophisticated
nonlinear solvers such as the simulated annealing method
\cite{NumericalRecipes}.

In this paper we investigate the possibility of making the CT
equations simpler and introduce some alternative forms which
follow from the analytical structure of the method. We shall
derive  equations  for calculating the unknown generalized angular
momenta, $L$'s, from the finite  input set of phase shifts, $\delta_l$'s. The
characteristic property of these equations is that they  do not
involve any complicated matrix inversion. The simplification is
only possible if the even angular momenta can separately be
handled from the odd ones. Physically such a situation occurs in
certain case of identical bosonic (fermionic) collision when only
even (odd) partial waves are contributing to the differential
cross section and, thus, only phase shifts belonging to even (odd)
angular momenta can be derived from experiments. We note that according to references \cite{Ramm,Horvath}, the even (odd) partial wave phase shifts alone would result in the same (unique) potential if an infinite subset of them would be used as input and a proper constructive procedure would be known. In case of finite set of phase shifts, however, when also even and odd partial waves contribute generally to the scattering, we can still use our simplified forms as approximations, and we shall provide various
approximate examples for such usages. Finally, we also investigate
the possibility of using  the so called one-term solution (or
total decoupling case) approximation $L=l-2\delta_\ell/\pi$, which
can be obtained from the CT equations with $N=1$.

In the next section  we collect all the equations necessary to outline the CT method. The simplified equations valid in case of even or odd angular momenta are derived in section 3 and the examples for the various applications with and without approximations are presented in section 4. Section 5 is devoted to the conclusion.

\section{Cox-Thompson method}
Let us introduce first the dimensionless quantities,  $x=kr$ for the distance, and $q(x)=\frac{V(x/k)}{E}$ for the potential where $E= {\hbar^2k^2}/{2m}$ denotes the scattering energy, $m$ the reduced mass, and $k$ the wave number.

The inverse potential corresponding to the radial scattering Schr\"odinger equation reads as follows
\begin{equation}\label{pot}
q(x)=-\frac{2}{x}\frac{ d}{ d x}\frac{K(x,x)}{x}
\end{equation}
where the quantity $K(x,x)$ is the diagonal part of the transformation kernel defined by a
Gel'fand-Levitan-Marchenko (GLM) type integral equation
\begin{equation}\label{GL}
K(x,y)=g(x,y)-\int_{0}^{x} d tt^{-2}K(x,t)g(t,y),\qquad x\geq y.
\end{equation}
In the CT method we take the following separable expansion for the input kernel:
\begin{equation}\label{CTang}
g(x,y)=\sum_{l\in S}\gamma_lu_l(x_<)v_l(x_>), \qquad \left\{\begin{array}{ll}
x_{<}\\
x_{>}
\end{array}\right\} =\left\{ \begin{array}{ll}
\min\\
\max
\end{array}\right\}(x,y)
\end{equation}
where $\gamma_l$ are expansion coefficients   (not required to be determined), $u_l\,(v_l)$ denote
the regular   (irregular) solutions of the free radial Schr\"odinger equation and the summation set $S$
consists of the physical angular momenta $\{l\}$.

To solve the GLM equation (\ref{GL}), Cox and Thompson \cite{Cox1970} introduce another separable
ansatz for the transformation kernel, in the form of a sum over an artificial angular momentum space $L\in T$
\begin{equation}\label{CTanK}
  K(x,y)=\sum_{L \in T}A_L(x)u_L(y)
\end{equation}
where $A_L$'s are unknown expansion functions and  $T$ is interpreted as a set of unknown shifted angular
momenta to be determined under the constraint that it has the same number $N$ of  {\it different} elements
as $S$ does, i.e., $N=|S|=|T|$ and $S\cap T=\emptyset$.

By inserting equations (\ref{CTang}) and (\ref{CTanK}) into the GLM equation (\ref{GL}) and using the linear
independence of the regular and irregular free solutions, one obtains equations for the determination of the
expansion functions $A_L(x)$ as follows
\begin{equation}\label{AL}
\sum_{L\in T}A_L(x)\frac{W[u_L(x),v_l(x)]}{l(l+1)-L(L+1)}=v_l(x)\qquad l\in S.
\end{equation}
Here, the only unknowns are the set $T$ with elements $L$, and $W$ denotes the Wronskian defined by $W[a,b]\equiv ab'-a'b$.

In order to determine the set $T$ one makes use of the Povzner-Levitan representation of the radial
 scattering wave function which reads as
\begin{equation}\label{PL}
\psi_l(x)=u_l(x)-\int_{0}^{x} d tt^{-2}K(x,t)u_l(t), \qquad l\in S
\end{equation}
whose asymptotic form containing the input phase shift data, takes the form
 \begin{equation}\label{PLa}
B_l\sin(x-l\frac{\pi}{2}+\delta_l)=\sin(x-l\frac{\pi}{2})-\sum_{L\in
T}A_L^{\rm{a}}(x)\frac{\sin((l-L)\frac{\pi}{2})}{l(l+1)-L(L+1)},\,
l\in S.\,
\end{equation}
Here, $B_l$ is a normalization constant and we have defined the
asymptotic expansion functions $A_L^{\rm{a}}(x)\equiv
A_L(x\to\infty)$ which can be calculated from the asymptotic
version of equation (\ref{AL}) which is given by
\begin{equation}\label{ALa}
\sum_{L\in T}A_L^{\rm{a}}(x)\frac{\cos((l-L)\frac{\pi}{2})}{l(l+1)-L(L+1)}=
-\cos(x-l\frac{\pi}{2}),\qquad l\in S.
\end{equation}
Using the last two equations, (\ref{PLa}) and (\ref{ALa}), one can easily derive  the following
 equations for the determination of the unknown $L$'s from the input phase shifts, $\delta_l$'s
 \cite{Apagyi2003,Melchert2006}:
\begin{equation}\label{general}
S_l=\frac{1+ i\mathcal{K}_l^+}{1- i\mathcal{K}_l^-}\qquad {\rm{or}}
\qquad\ \tan(\delta_l)=\frac{\mathcal{K}_l^++\mathcal{K}_l^-}{2+ i(\mathcal{K}_l^+-\mathcal{K}_l^-)},
\qquad l\in S
\end{equation}
where $S_l= e^{2 i\delta_l}$ and the "shifted" reactance matrix elements are defined as
\begin{equation}
\mathcal{K}_l^{\pm}=\sum_{L\in T, l'\in S}[M_{\sin}]_{lL}[M_{\cos}^{-1}]_{Ll'} e^{\pm  i(l-l')\pi/2},
\qquad l\in S,
\end{equation}
with
\begin{equation}\label{Msincos}
\left\{\begin{array}{ll}
M_{\sin}\\
M_{\cos}
\end{array}\right\}_{lL} = \frac{1}{L(L+1)-l(l+1)}\left\{ \begin{array}{ll}
\sin\left((l-L)\frac{\pi}{2}\right)\\
\cos\left((l-L)\frac{\pi}{2}\right)
\end{array} \right\},\qquad l\in S,\,L\in T.
\end{equation}
The appearance of the inverse of the matrix $M_{\cos}$ containing the unknown $L$'s is the
problem that makes the solution of the nonlinear equations (\ref{general}) especially hard.

If one succeeds to solve either one or both of the highly nonlinear equations (\ref{general})
and thus finds the set $T$, the CT inverse potential $q_{\rm{CT}}(x)$ can   easily be obtained
by employing equations (\ref{AL}), (\ref{CTanK}), and (\ref{pot}).

\section{Simplified solutions and approximations}

In this section we present simplifications to equations (\ref{general}) which can be used if only
even (odd) partial waves are arising during the collision. Otherwise the simplified equations can
be employed to construct different approximations which will be discussed in separate subsections.

\subsection{Even (odd) angular mom\-ent\-um treat\-ment }
Let us differentiate equation (\ref{ALa})  twice with respect to the variable $x$. Then we arrive
at the following equation
\begin{equation}\label{difA}
\frac{ d^2A_L^{\rm{a}}(x)}{ d x^2}=-A_L^{\rm{a}}(x),
\end{equation}
which has a periodic
  solution as
\begin{equation}\label{difAsol}
A_L^{\rm{a}}(x)=a_L\cos(x)+b_L\sin(x).
\end{equation}
 Now, by inserting this solution (\ref{difAsol}) into equation (\ref{ALa}) and taking into account
 the independence of the sine and cosine functions, one gets the following two equations
\begin{equation}\label{albl}
\sum_{L\in T}\left\{\begin{array}{ll}
a_L\\
b_L
\end{array}\right\}\frac{\cos\left((l-L)\frac{\pi}{2}\right)}{L(L+1)-l(l+1)} = \left\{\begin{array}{ll}
\cos\left(l\frac{\pi}{2}\right)\\
\sin\left(l\frac{\pi}{2}\right)
\end{array} \right\},\qquad l\in S.
\end{equation}
Consider the decomposition
$$
S=S_{\rm{e}}\cup S_{\rm{o}}
$$
where $S_{\rm{e}}$ and $S_{\rm{o}}$ contains, respectively, the
even and odd elements of $S$. Instead of (14) we consider two
systems:
\begin{equation*}
\sum_{L\in T_{\rm{e}}}\left\{\begin{array}{ll}
a_L\\
b_L
\end{array}\right\}\frac{\cos\left((l-L)\frac{\pi}{2}\right)}{L(L+1)-l(l+1)} = \left\{\begin{array}{ll}
\cos\left(l\frac{\pi}{2}\right)\\
\sin\left(l\frac{\pi}{2}\right)
\end{array} \right\},\qquad l\in S_{\rm{e}}
\end{equation*}
and
\begin{equation*}
\sum_{L\in T_{\rm{o}}}\left\{\begin{array}{ll}
a_L\\
b_L
\end{array}\right\}\frac{\cos\left((l-L)\frac{\pi}{2}\right)}{L(L+1)-l(l+1)} = \left\{\begin{array}{ll}
\cos\left(l\frac{\pi}{2}\right)\\
\sin\left(l\frac{\pi}{2}\right)
\end{array} \right\},\qquad l\in S_{\rm{o}},
\end{equation*}
where $|T_{\rm{e}}|=|S_{\rm{e}}|$, $|T_{\rm{o}}|=|S_{\rm{o}}|$ and
$T_{\rm{e}}\cap S_{\rm{e}}=\emptyset$, $T_{\rm{o}}\cap
S_{\rm{o}}=\emptyset$. These systems have the solutions
\begin{equation}\label{al}
a_L=\frac{\prod_{l\in S_{\rm{e}}}(L(L+1)-l(l+1))}{\prod_{L'\in T_{\rm{e}}\backslash\{L\}}
(L(L+1)-L'(L'+1))}\frac{1}{\cos\left(L\frac{\pi}{2}\right)},\qquad b_L=0,\qquad L\in T_{\rm{e}},
\end{equation}
and
\begin{equation}\label{bl}
a_L=0,\qquad b_L=\frac{\prod_{l\in S_{\rm{o}}}(L(L+1)-l(l+1))}{\prod_{L'\in T_{\rm{o}}\backslash\{L\}}
(L(L+1)-L'(L'+1))}\frac{1}{\sin\left(L\frac{\pi}{2}\right)},\qquad L\in T_{\rm{o}},
\end{equation}
respectively. In case of $T_{\rm{e}}\cap T_{\rm{o}}\ne\emptyset$
the formulae (\ref{al}) and (\ref{bl}) may assign different values
to the same  $a_L$ and $b_L$ but this is not a real ambiguity
because we always use separately the solution vectors (\ref{al})
and (\ref{bl}) in our calculations.

Now, by using the explicit expressions (\ref{al}) in equations
(\ref{difAsol})  and (\ref{PLa}), one obtains the final solution
to the CT method as
 \begin{equation}\label{even}
\tan(\delta_{l})=-\sum_{L \in T_{\rm{e}}}\frac{\prod_{l'\in
S_{\rm{e}}\backslash\{l\}}(L(L+1)-l'(l'+1))}{\prod_{L'\in
T_{\rm{e}}\backslash\{L\}}(L(L+1)-L'(L'+1))} \tan\left(L\frac{\pi}{2}\right),\qquad
l\in S_{\rm{e}},
\end{equation}
 valid for the case of even $l$'s.
Similarly, using equations (\ref{bl}) we get the solution to the CT method as
\begin{equation}\label{odd}
\tan(\delta_{l})=\sum_{L \in T_{\rm{o}}}\frac{\prod_{l'\in
S_{\rm{o}}\backslash\{l\}}(L(L+1)-l'(l'+1))}{\prod_{L'\in
T_{\rm{o}}\backslash\{L\}}(L(L+1)-L'(L'+1))} \cot\left(L\frac{\pi}{2}\right),\qquad
l\in S_{\rm{o}}
\end{equation}
which are valid in the case of odd $l$'s. These equations    determine the unknown sets $T_{\rm{e}}$
or $T_{\rm{o}}$ of shifted angular momenta $L$ and, thus, replace either of the corresponding
generic solutions (\ref{general}).

Notice the simplified structure of the nonlinear equations  (\ref{even})  and (\ref{odd}),
compared to equations (\ref{general}). While equations  (\ref{general})   contain an explicit
matrix inversion of a matrix involving the unknowns of shifted angular momenta, $L$'s, formulas
(\ref{even})  and (\ref{odd}) do not require such a nonlinear operation.  They
'only' contain products and the tangent (cotangent) operations and are thus presumably easier
to be solved for the sets $T_{\rm{e}}$ or $T_{\rm{o}}$, if the respective input phase shifts are given.

Finding the sets $T_{\rm{e}}$ or $T_{\rm{o}}$, the corresponding potentials $q_{\rm{e}}(x)$
or $q_{\rm{o}}(x)$ can be obtained similarly as in the general case, by employing equations
(\ref{AL}), (\ref{CTanK}), and (\ref{pot}).

\subsection{Equivalence of solutions (\ref{general}) and (\ref{even}) or (\ref{odd})}

By explicit calculation one can check the equivalence of equations
(\ref{general}) and (\ref{even}) or (\ref{odd}) for the special
cases of even or odd $l$'s. For the case of either even or odd
$l$'s, the relation $\mathcal{K}_l^+=\mathcal{K}_l^-$ holds. Now,
specifying ourselves to the even $l$ case only, $l\in S_{\rm{e}}$,
the second of the general solution (\ref{general}) can be written
as
\begin{equation}\label{evengeneral}
\tan(\delta_l)=\sum_{L\in T_{\rm e}, l'\in S_{\rm
e}}[M_{\sin}]_{lL}[M_{\cos}^{-1}]_{Ll'}(-)^{ (l-l') /2},\qquad
l\in S_{\rm e}.
\end{equation}
Using equations (\ref{ALa}),   and (\ref{difAsol}), we get the
expression
\begin{equation}\label{20}
a_L\cos(x)=\sum_{l'\in S_{\rm{e}}}[M_{\cos}^{-1}]_{Ll'}\cos(x-l'{\pi\over 2}),\qquad L\in{T_{\rm{e}}}
\end{equation}
which simplifies to
\begin{equation}\label{21}
a_L =\sum_{l'\in S_{\rm{e}}}[M_{\cos}^{-1}]_{Ll'}(-)^{l'/2},\qquad L\in{T_{\rm{e}}}.
\end{equation}
By multiplying both sides of this equation by $(-)^{l/2}[M_{\sin}]_{lL}$, and performing the sum
 over $L$'s, one may write
\begin{equation}\label{22}
\sum_{L\in T_{\rm{e}}}a_L[M_{\sin}]_{lL} (-)^{l/2}=\sum_{L\in
T_{\rm{e}},l'\in
S_{\rm{e}}}[M_{\sin}]_{lL}[M_{\cos}^{-1}]_{Ll'}(-)^{(l-l')/2},\quad
l\in{S_{\rm{e}}}.
\end{equation}

According to equation (\ref{evengeneral}), the right hand side is
already equal to $\tan(\delta_l)$, and, by noting that the matrix
$M_{\sin}$ on the left hand side can be written, on account of
(\ref{Msincos}), as $[M_{\sin}]_{lL}=-(-)^{l/2}\sin(L{\pi\over
2})/(L(L+1)-l(l+1))$, one gets the formula
\begin{equation}\label{23}
\tan(\delta_l)=-\sum_{L\in T_{\rm{e}}}a_L{\sin(L{\pi\over 2})\over {L(L+1)-l(l+1)}},\qquad l\in{S_{\rm{e}}}
\end{equation}
which is the same as equation (\ref{even}) if one takes into consideration the solution (\ref{al}) for
the coefficient $a_L, L\in T_{\rm{e}}$.

A similar procedure can be applied to proving equivalence of equations (\ref{general}) and (\ref{odd})
 for odd $l$'s

\subsection{Approximations}

Equations (\ref{even}) and (\ref{odd}) can be used to derive inverse potentials only in the case when
either even or odd partial waves are arising during the collision process. In other words, only identical
bosonic or fermionic scattering can be treated by the simplified equations (\ref{even}) and (\ref{odd}).
However, the simplified equations offer several possibilities to introduce various approximate treatments
of the general scattering case. The type of approximations will be classified according to the level it
is applied to.

\subsubsection{Potential approximation, A}

If the general equations (\ref{general}) can be solved by neither of the nonlinear solvers at hand,
one may try to assess the inverse potential by solving the simplified equations (\ref{even}) and
(\ref{odd}) for the sets  $T_{\rm{e}}$ and $T_{\rm{o}}$. Then, separately constructing the
corresponding potentials $q_{\rm{e}}(x)$ and $q_{\rm{o}}(x)$, one simply adds them together
to get an approximation of the interaction potential, $q_{\rm{A}}(x)=q_{\rm{e}}(x)+q_{\rm{o}}(x)$.

\subsubsection{T-set approximation,  T }

One may try to approximate the set of the shifted angular momenta themselves. By unifying the
two sets obtained by solving equations (\ref{even}) and (\ref{odd}), one gets the T-set
approximation $T_{\rm{a}}=T_{\rm{e}}\cup T_{\rm{o}}$. Using this approximate set $T_{\rm{a}}$
in conjunction with equations (\ref{AL}), (\ref{CTanK}), and (\ref{pot}), one gets the approximate
inverse potential $q_{\rm{T}}(x)$.

\subsubsection{One-term approximation,   L }

If the collision is dominated overwhelmingly by a single partial wave (as in the case of resonance
scattering) then the equations (\ref{general}) are to be solved  at $N=1$, and this results in
the simple expression $L=l-2\delta_l/\pi$ for the shifted angular momentum, assuming that the
$l$th partial wave is dominating.
If however this is not the case, one  still may try to use the approximate expressions
 \begin{equation}\label{L_a}
L_{\rm{a}}=l-2\delta_l/\pi
  \end{equation}
to form an approximate set $T_{\rm{L}}$.
Using this approximate set $T_{\rm{L}}$ in conjunction with equations (\ref{AL}), (\ref{CTanK}),
and (\ref{pot}), one gets the approximate inverse potential denoted by $q_{\rm{L}}(x)$.

\section{Examples}
In this section we first apply the simplified solution (\ref{even}) to calculate an effective
 potential related to phase shifts derived from  bosonic collisions. Then, two exploratory calculations
 follow demonstrating applicability of different approximations introduced in the preceding section.

\subsection{ Effective $^{87}$Rb $+$ $^{87}$Rb potential at $E=303$ $\mu$K}

In this subsection we recalculate one result of \cite{Schumayer2008}
where effective Rb-Rb inter-atomic potentials have been derived from ultracold Bose-gas collision data.

The inverse calculation is performed by using equation
(\ref{even}) at  $E=303$ $\mu$K. The corresponding phase shifts
$\delta_l^{\rm{orig}}$ measured (in rad) are as follows:
$-1.287,\,1.635,\,0.005$  related to partial waves with $l=0,2,4$,
respectively.

\begin{figure}[ht]
\begin{center}
  \includegraphics[width=11cm]{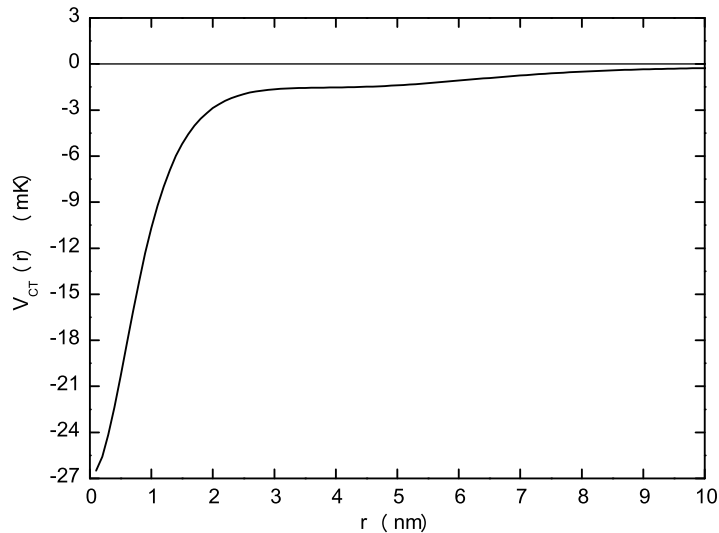}\\
  \caption{Effective $^{87}$Rb$+$$^{87}$Rb inverse potential $V_{\rm{CT}}$ (in mK) as a function
  of the radial distance $r$ (in nm) at c.m. energy $E =303\,\mu$K.}\label{Rb}
\end{center}
\end{figure}

The resulted inverse potential $V_{\rm{CT}}(r)$ is plotted in
figure \ref{Rb}. To control the procedure we recalculated the
phase shifts from the  inverse potential and the difference
$\Delta^{\rm{CT}}=|\delta_l^{\rm{orig}}-\delta_l^{\rm{CT}}|$
between the phase shifts are as follows: $0.014,\,0.042,\,0.004$
for $l=0,2,4$, respectively.

The result shown in  figure \ref{Rb} is identical with that obtained by \cite{Schumayer2008} using
the generic CT procedure. Indeed we also calculated the inverse potential by using equation
(\ref{general}), and obtained the same result as depicted in figure \ref{Rb}. In other words, the equivalence
 of equation (\ref{even}) and (\ref{general}) in case of even  $l$'s has been numerically verified too.

\subsection{Gauss potential}

As a next example we explore the applicability of different approximations introduced in subsection 3.2.
For the analysis we use the prescribed potential of Gauss-form:
\begin{equation}\label{Gauss}
V_{\rm{G}}(r)=-2\exp{(-5r^2)}
\end{equation}
where both distance $r$ and energy are measured in atomic units (au).

\begin{figure}[ht]
\begin{center}
 \includegraphics[width=11cm]{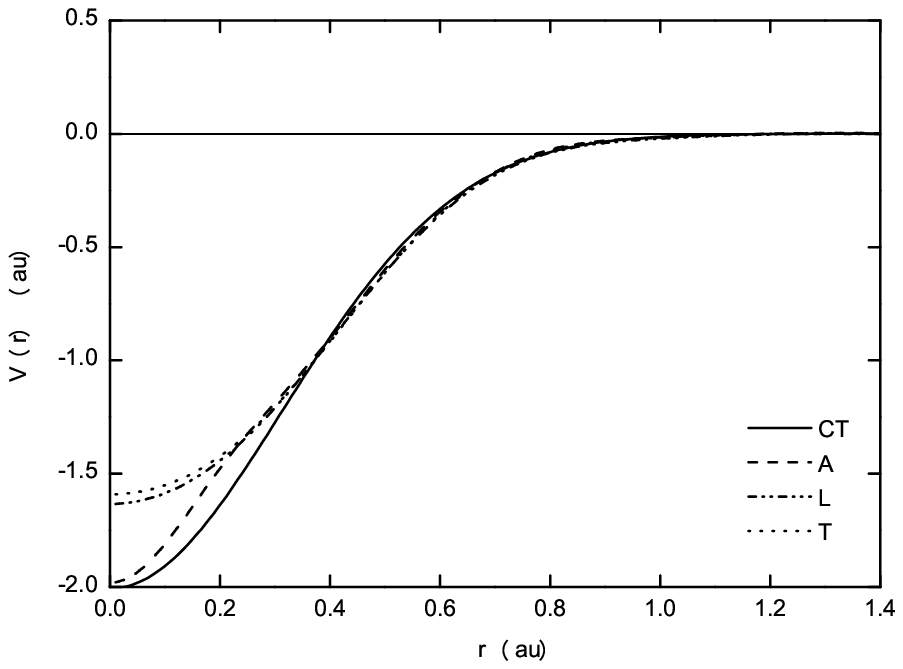}
  \caption{Inverse potentials $V(r)$ obtained from input phase shifts  $\delta_l^{\rm{orig}}$ [see table \ref{tab1}]  as a function of the radial distance $r$   at  energy $E=18$ ($k=6$) au. Curves  obtained  by  the CT,  and  approximate methods are labeled according to the procedures discussed in the text.}
\label{Gaussfig}
\end{center}
\end{figure}

Potential (\ref{Gauss}) provides a set of input phase shifts $\delta_l^{\rm{orig}}$ listed in table \ref{tab1}
 at  scattering energy of $E=18$ au ($k=6$ au). Using this input set one calculates the CT inverse potential
 $V_{\rm{CT}}(r)=E \,q_{\rm{CT}}(kr)$ which is depicted in figure \ref{Gaussfig}. Note that $V_{\rm{CT}}$ is the same as $V_{\rm{G}}$, within the width of line, therefore plot of $V_{\rm{G}}$ is omitted in figure \ref{Gaussfig}.

\begin{table}
\caption{\label{tab1} Original phase shifts $\delta_l^{\rm{orig}}$ produced by the Gauss potential (\ref{Gauss}) at scattering energy $E=18$ au. Shifted angular momenta $L$ corresponding to solution of equations (\ref{general}), (\ref{even}), or (\ref{odd}), and (\ref{L_a}).}

\begin{center}
\begin{tabular}{ccccc}
\hline
$\phantom{0}l$ &  $\delta_l^{\rm{orig}}$ & \phantom{0}(\ref{general})  & (\ref{even}) or (\ref{odd}) & \phantom{0}(\ref{L_a})  \\
\hline
   \phantom{0}0 &       0.1294 & --0.0893  & \phantom{0}\phantom{0}--0.0809 & --0.0824    \\
   \phantom{0}1 &       0.0964 &  \phantom{0}0.9392  &  \phantom{0}\phantom{0}\phantom{0}0.9391 &  \phantom{0}0.9386    \\
   \phantom{0}2 &       0.0535 &  \phantom{0}1.9676  &  \phantom{0}\phantom{0}\phantom{0}1.9666 &  \phantom{0}1.9659    \\
   \phantom{0}3 &       0.0232 &  \phantom{0}2.9865  &  \phantom{0}\phantom{0}\phantom{0}2.9861 &  \phantom{0}2.9852    \\
   \phantom{0}4 &       0.0082 &  \phantom{0}3.9955  &  \phantom{0}\phantom{0}\phantom{0}3.9954 &  \phantom{0}3.9948    \\
   \phantom{0}5 &       0.0025  &  \phantom{0}4.9989  &  \phantom{0}\phantom{0}\phantom{0}4.9989 &  \phantom{0}4.9984    \\
   \phantom{0}6 &       0.0006  &  \phantom{0}5.9999  &  \phantom{0}\phantom{0}\phantom{0}5.9999 &  \phantom{0}5.9996    \\
   \phantom{0}7 &      0.0001  &  \phantom{0}7.0001  &  \phantom{0}\phantom{0}\phantom{0}7.0001 &  \phantom{0}6.9999   \\
   \phantom{0}8 &      0.0000  & \phantom{0}8.0002  &  \phantom{0}\phantom{0}\phantom{0}8.0002 &  \phantom{0}8.0000    \\
   \phantom{0}9 &     0.0000   &  \phantom{0}9.0001  &  \phantom{0}\phantom{0}\phantom{0}9.0001 &  \phantom{0}9.0000  \\
    10 &     0.0000  & 10.0001  & \phantom{0}\phantom{0}10.0001 & 10.0000    \\
\hline
\end{tabular}
\end{center}
\end{table}

The different approximations  $V_{\rm{A}}$, $V_{\rm{L}}$, and $V_{\rm{T}}$ can thus be compared to  $V_{\rm{CT}}$ which may be taken to be exact. In figure \ref{Gaussfig} we see that, for this particular example, the potential $V_{\rm{A}}(r)=E\,q_{\rm{A}}(kr)$ is a good approximation to the original one in view of its initial behaviour (depth) and asymptotical property (range). Recall that approximation $V_{\rm{A}}$ is obtained by simply adding inversion approximations $V_{\rm{e}}$ and $V_{\rm{o}}$ derived by inverting the separate sets $S_{\rm{e}}$ and $S_{\rm{o}}$ of phase shifts belonging, respectively, to the sets of even and odd angular momenta $l'$s. Approximation $V_{\rm{T}}$ is obtained by unifying the calculated sets $T_{\rm{e}}$ and $T_{\rm{o}}$ of shifted angular momenta, $L'$s,   listed in table \ref{tab1} under heading (\ref{even}) or (\ref{odd}). Finally, the approximation $V_{\rm{L}}$ has been obtained by simply using the one-term approximate  values $L_{\rm{a}}$, equation (\ref{L_a}), which are also listed in table \ref{tab1} under heading (\ref{L_a}). It is interesting to see in figure \ref{Gaussfig} that this (totally analytical) version, the approximation $V_{\rm{L}}$ provides a somewhat better result than that of the more involved approximation $V_{\rm{T}}$.

\subsection{$n +\, ^{12}$C scattering at $E=10$ MeV}
Our last example is taken from nuclear physics. Chen and Thornow \cite{Chen2005}
have derived 88 sets of complex-valued phase shifts  from a comprehensive analysis of $\rm{n}$ scattering by $^{12}$C target nucleus in the scattering energy region 7 MeV $\le E_{\rm{n}}^{\rm{lab}}\le$ 24 MeV. Because of the spin-orbit coupling, each partial wave provides two phase shifts, $\delta_l^+$ and $\delta_l^-$.   In case of weak spin-orbit coupling the combined phase shift $\delta_l=[(l+1)\delta_l^+ +l\delta_l^-]/(2l+1)$ are characteristic of the underlying central potential \cite{Leeb1995}.

\begin{figure}[ht]
\begin{center}
\rotatebox{0}{
 \includegraphics[width=11cm]{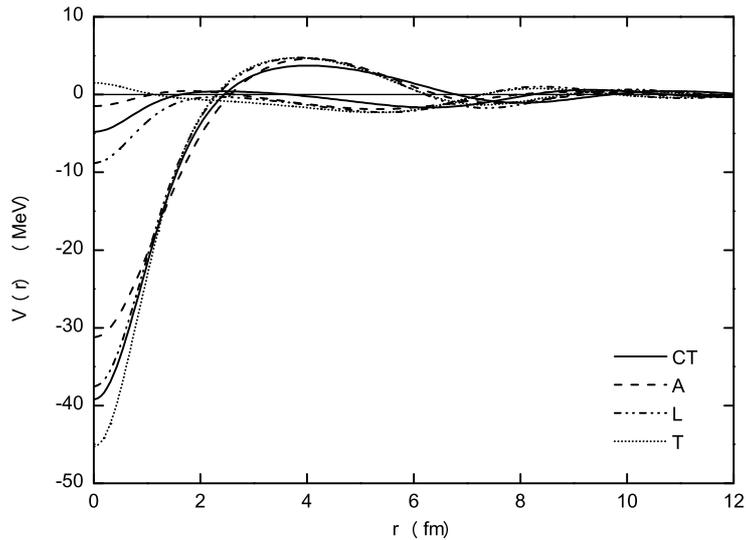}
}
   \caption{Inverse potentials $V(r)$ obtained from input phase shifts  $\delta_l^{\rm{orig}}$  and elasticities  $\eta_l^{\rm{orig}}$  [see table \ref{tab2}]  as a function of the radial distance $r$   at  energy $E_{\rm{n}}^{\rm{lab}}=10$ MeV ($E_{\rm{n}}^{\rm{c.m}}=9.23$ MeV, $k=0.638$ fm$^{-1}$). Curves  obtained  by  the CT,  and  approximate methods are labeled according to the procedures discussed in the text. The imaginary part of the potentials concentrates around the abscissa, meaning a weak absorption.}
\label{fig3}
\end{center}
\end{figure}

One set of such combined phase shifts,  valid to the neutron scattering by  $^{12}$C at the energy of $E_{\rm{n}}^{\rm{lab}}=10$ MeV is listed in table \ref{tab2}. Here, $\delta_l^{\rm{orig}}$ denotes the real part of the combined phase shifts, $\rm{Re}{\delta_l}$, and $\eta_l^{\rm{orig}}$ stands for the elasticity, $\eta_l^{\rm{orig}}=|\exp(2 i\delta_l)|$. The results of the various inversion procedures are shown in figure \ref{fig3}.

\begin{table}
\caption{\label{tab2} Original phase shifts $\delta_l^{\rm{orig}}$ and elasticities $\eta_l^{\rm{orig}}$ taken from the phase shift data of \cite{Chen2005} for  $\rm{n}$ scattering by $^{12}$C nucleus at the energy $E_{\rm{n}}^{\rm{lab}}=10$ MeV. Real (left column) and imaginary (right column) part of the shifted angular momenta $L$ listed under headings CT, A, and L, respectively, correspond to solutions of equations (\ref{general}), (\ref{even}), or (\ref{odd}), and (\ref{L_a}).}

\begin{center}
\begin{tabular}{ccccccccc}
\hline
$l$ &  $\phantom{-}\delta_l^{\rm{orig}}$ & $\eta_l^{\rm{orig}}$ & \phantom{-} CT  & \phantom{-} CT  &  \phantom{-} A & \phantom{-} A  & \phantom{-} L & \phantom{-} L  \\
\hline
    0  & \phantom{-}$0.827$  & $0.580$ & $-0.581$ & $-0.085$ & $-0.583$  & $-0.076$ & $-0.527$  & $-0.173$ \\
    1  & $-0.562$ & $0.723$ & \phantom{-}$1.226$  &  \phantom{-}$0.001$ & \phantom{-}$1.360 $  & $-0.122$ & \phantom{-}$1.358$ & $-0.103$   \\
    2  & $-0.365$ & $0.526$ & \phantom{-}$2.349$  & $-0.192$ &  \phantom{-}$2.259 $ & $-0.203$ &  \phantom{-}$2.232$ & $-0.205$   \\
    3  & \phantom{-}$0.057$ & $0.846$  & \phantom{-}$2.981$  & $-0.073$ &  \phantom{-}$3.011 $ & $-0.084$ &  \phantom{-}$2.964$ & $-0.053$   \\
    4  & \phantom{-}$0.021$ & $0.959$  & \phantom{-}$4.010$  & $-0.062$ &  \phantom{-}$4.001 $ & $-0.050$ &  \phantom{-}$3.987$ & $-0.013$   \\
\hline
\end{tabular}
\end{center}
\end{table}

In order  to draw conclusion about the applicability of the
different approximate inverse procedures A, T, and L, one can
compare the corresponding potential curves to the one obtained by
the CT method which is assumed the best. (There is no model
potential in this case.)  Indeed, if one calculates back the phase
shifts and elasticities that the different inverse potentials
$V_{\rm{CT}}$, $V_{\rm{A}}$, $V_{\rm{T}}$, and $V_{\rm{L}}$ are
providing, we see that the reproduction is the best for the CT
potential. All these facts can be studied in table \ref{tab3}
where we list the differences $\Delta_l$ and $\Xi_l$ between the
original data $\delta_l^{\rm{orig}}$ and $\eta_l^{\rm{orig}}$ and
the re-calculated ones provided  by the inverse potentials of the
respective  methods. The  potential curves $V_{\rm{CT}}$,
$V_{\rm{A}}$, $V_{\rm{T}}$, and $V_{\rm{L}}$ are drawn in figure
\ref{fig3} and we see that the curves obtained by the methods A,
T, and L approximate well the inverse potential $V_{\rm{CT}}$
considered to be the reference. Because the numerical reproduction
is also surprisingly good   we may conclude that the proposed
approximations to the CT method can be used for a global
orientation about the nature of the underlying interaction, if
input phase shifts are known from another source and the solution
of the generic equation (\ref{general}) is not possible.

\begin{table}
\caption{\label{tab3} Differences $\Delta_l$ and $\Xi_l$ between the
original data and the calculated ones provided by the inverse
potentials (shown in figure \ref{fig3}) obtained by the different
inverse methods CT, A, T, and L.}

\begin{center}
\begin{tabular}{ccccccccc}
\hline
$l$ & $\phantom{0}\Delta_l^{\rm{CT}}$&$\phantom{0}\Xi_l^{\rm{CT}}$& $\phantom{0}\Delta_l^{\rm{A}}$&$\phantom{0}\Xi_l^{\rm{A}}$& $\phantom{0}\Delta_l^{\rm{L}}$&$\phantom{0}\Xi_l^{\rm{L}}$& $\phantom{0}\Delta_l^{\rm{T}}$&$\phantom{0}\Xi_l^{\rm{T}}$\\
\hline
$0$  &$0.014$&$0.004$&$0.027$&$0.054$&$0.027$&$0.111$&$0.057$&$0.023$\\
$1$ & $0.015$&$0.022$&$0.085$&$0.085$&$0.120$&$0.276$&$0.121$&$0.315$\\
$2$ & $0.040$&$0.054$&$0.122$&$0.093$&$0.158$&$0.099$&$0.127$&$0.103$\\
$3$  &$0.029$&$0.071$&$0.025$&$0.060$&$0.070$&$0.146$&$0.009$&$0.119$\\
$4$  &$0.036$&$0.056$&$0.036$&$0.013$&$0.000$&$0.001$&$0.018$&$0.060$\\
\hline
\end{tabular}
\end{center}
\end{table}

\section{Conclusion}
By observing the simple analytical property   (\ref{difAsol})  of
the asymptotic expansion functions $A_L^{\rm{a}}(x)$ of the
transformation kernel $K(x,y)$ involved in the Cox-Thompson (CT)
inverse scattering method at fixed energy, simplified solutions of
the CT method have been derived [see equations (\ref{even}) and
(\ref{odd})] applicable to certain special scattering of identical
bosons (fermions) when only even (odd) partial waves are arising
(and thus measurable).

The new formulas (\ref{even}) and (\ref{odd}) obtained are easier to be solved by usual nonlinear solvers (such as that based, e.g., on the Newton-Raphson procedure \cite{NumericalRecipes}). This is because the new formulas do not involve the inverse of a matrix containing the unknowns themselves; they 'only' contain product and tangent (cotangent) operations. Therefore the range of applicability of the new formulas is wider than that of the generic equation (\ref{general}).  However,  identical bosonic (fermionic) collision experiments are rare in practice. Therefore we have developed several approximations too, based on the new formulas (\ref{even}) and (\ref{odd}), classified according to the level in which they have been introduced. By taking examples from atomic collision experiment, potential scattering and nuclear physics, we have demonstrated the wide applicability of the new equations
which make the solution of the CT inverse scattering method at fixed energy easier.

\section*{Acknowledgements}
One of us (T. P.) wishes to express his gratitude to
Professor Werner Scheid for kind hospitality and reading the
manuscript. This work was supported by the Hungarian Scientific
Research Fund, under contracts OTKA-T47035, T61311, IN67371 and
the MTA-DFG grant (436 UNG 113/158).

\end{document}